\documentstyle[12pt]{article}
\begin{document}
\begin{titlepage}
\begin{center}

{\Large\bf MIGHT FAST B-VIOLATING TRANSITIONS BE FOUND SOON?}

\vskip 1.5cm

{\large\bf Vadim A. Kuzmin}
\vskip 0.8cm
{\it Institute for Nuclear Research of Russian Academy of Sciences},\\
{\it 60th October Anniversary Prosp. 7a, Moscow 117312, Russia}
{\footnote[1]{\it Permanent address. E-mail : kuzmin@ms2.inr.ac.ru .}}

and 

{\it Max-Planck-Institut fuer Physik,} \\
{\it Foeringer Ring 6,}\\
{\it 80805 Muenchen, Germany}
\vskip 0.5cm    
\end{center}
\vskip 2cm

\begin{abstract}
We claim that there might exist a new interaction leading to very fast 
baryon-number violating processes quite observable in the laboratory 
conditions, provided all three generations are simultaneously involved.   
\end{abstract}
\endcenter
\vskip 3cm
\begin{center}
{\it Talk presented at the International Workshop on Future Prospects of 
Baryon Instability Search in p-Decay and $n-{\bar n}$ Oscillation Experiments, 
Oak Ridge, Tennessee, March 28-30 1996.}
\end{center}
\end{titlepage}

{\it \bf 1. Baryon-Antibaryon Mixings and Oscillations}.

It was conjectured few years ago ${\cite {kuz1}}$ that there could take place 
very fast baryon-number violating transitions with quite observable rates 
which however escaped observation up to now due to some specifics of the 
interaction. Specifically, it was conjectured ${\cite {kuz1}}$ that there 
could be a baryon-number violating coupling originating in particles of 
electroweak masses and coupling strengths, provided all three generations are 
simultaneously involved. 
An example of what I am talking about is given by the coupling 
\begin{equation} 
\epsilon_{ijk} \lambda \phi_{i} q_{j}q_{k} , 
\end{equation}
\noindent 
$\phi$ being color-triplet scalar fields, $q$ being right-handed quark 
fields, $\lambda$ being the coupling constant and $i,j,k =1,2,3$ 
are family indices. Note that the corresponding scalar fields are quite 
generic in the context of any GUT.    
Being interested in various aspects of baryon-
antibaryon mixings and oscillations, I exhibited special interest in the 
$\Xi_{b}$ (bus) baryon as the lightest one composed of quarks of all three 
generations which might undergo a lot of mixing with its antiparticle ( see 
Fig. 1). 

\newcounter{cms}
\setlength{\unitlength}{1mm}
\begin{center}
\begin{picture}(100,70)(0,0)
\put(34,9){\vector(0,-1){16}}
\put(66,9){\vector(0,-1){16}}
\put(34,9){\vector(-1,0){16}}
\put(66,9){\vector(1,0){16}}
\put(50,44){\vector(-1,1){12}}
\put(50,44){\vector(1,1){12}}
\put(52,35){B}
\put(38,20){U}
\put(62,20){S}
\put(39,49){s}
\put(61,49){u}
\put(25,11){s}
\put(75,11){u}
\put(36,0){b}
\put(62,0){b}
\put(50,25){\line(0,1){5}} 
\put(50,32){\line(0,1){5}}
\put(50,39){\line(0,1){5}}
\put(50,25){\line(1,-1){4}}
\put(56,19){\line(1,-1){4}}
\put(62,13){\line(1,-1){4}}
\put(50,25){\line(-1,-1){4}}
\put(44,19){\line(-1,-1){4}}
\put(38,13){\line(-1,-1){4}}
\end{picture}
\vspace{20mm}

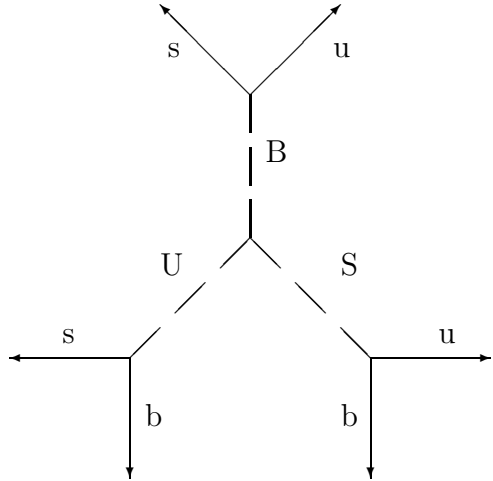
Fig. 1. $(bus)-({\bar b}{\bar u}{\bar s})$ coupling. The dashed lines are 
color-triplet scalar fields $\phi$; the solid lines are right-handed 
quark fields. 
\end{center}
\vspace{1.5cm}
       
By rapid transitions $(bus) \leftrightarrow ({\bar b}{\bar u}{\bar s})$ I mean 
that the transition time, 
$\tau_{bus \leftrightarrow {\bar b}{\bar u}{\bar s}}$, is not excluded to be 
of order of the weak decay lifetime,  
\begin{equation} 
\tau_{bus \leftrightarrow {\bar b}{\bar u}{\bar s}} \sim 10^{-13} s .
\end{equation}

It does not seem that such couplings would result in problems with FCNC and/or 
hyperon $\leftrightarrow$ anti-hyperon transitions.  
It seems that the most stringent constraints on the magnitude of the coupling 
under consideration come from results of experimental searches of matter 
instability (proton decay, neutron-antineutron transitions in nuclei and in 
vacuum). However, what I am going to conjecture now, is the following. 
Remarkably enough, neutron-antineutron transitions originating from radiative 
electroweak corrections to the proposed interaction Fig. 1 being tremendously 
suppressed in comparison with 
$(bus) \leftrightarrow ({\bar b}{\bar u}{\bar s})$ transitions by factor 
$ \sim (G_{F}^2)^2 \sim 10^{-20}$ might be well in the right range for 
experimental searches, with 
\begin{equation} 
\tau_{n \leftrightarrow {\bar n}} \sim 10^{7}-10^{8} s.  
\end{equation}

Thus, fast enough baryon-number violating transitions might be looked for 
both by investigation of wrong signature weak decays of (bus)-like baryons and 
by searches of $n{\bar n}$ transitions.

{\it \bf 2. A Speculation on ALEPH Events}.

One may speculate on a possible relation of the assumed existence of these new 
scalars $\phi$ mediating baryon-number violating transitions to presumably 
observed recently 4-jet events by ALEPH. Such features of ALEPH events 
as no missing energy, absence of b-quarks in jets, relatively large yield 
in comparison with expectations might be easily understood in the framework of 
our hypothesis. Indeed, our colored scalars $\phi$  are coupled to quarks 
only and not to leptons. Second, if a pair of produced particles (with 
masses 55 GeV), giving two jets after their decay, is assumed to be a pair 
of $\phi$'s with the family index $j=3$, then one should not expect 
$b$-quarks in jets at all. Finally, if there is indeed some excess in number 
of events observed, it might be explained by large electric charge 
of some of $\phi$'s.   
    
{\it \bf 3. Conclusions.}

New rich physics might be well quite nearby! Searches are 
worthwhile both at accelerators and in low energy experiments and  
may proceed in several directions. Among them are, obviously, the following 
ones.

1. Production of pairs  $\phi{\bar \phi}$ at hadron and electron colliders 
(in experiments like ALEPH).

2. Production of (bus)-like baryon-antibaryon pairs, say, at $e^{+}e^{-}$ 
colliders, and search for wrong signature decays of produced baryons and 
antibaryons. Wrong signature is due to baryon-antibaryon mixing.

3. Search for $n{\bar n}$ oscillations in free neutrons beam. 
It might be well that even present sensitivity is already almost sufficient 
and we are close to observation of this phenomenon.  

4. Search for induced matter instability. 

The proposed new interaction might have a deep impact on generation of the 
baryon asymmetry of the Universe.
 
{\it \bf Acknowledgements.} 

I am grateful to J.D. Bjorken, S. Pokorski, L. Stodolsky and V. Zakharov  
for many encouraging discussions and comments and  I. Tkachev for 
the  stimulating interest  and enlightening discussions. 
I am thankful to L. Stodolsky for the interest in the work, support and warm 
hospitality at Max-Planck Institut fuer Physik, Muenchen, and 
Yu. Kamyshkov for his interest, enthusiasm and hospitality 
extended to me in Oak Ridge.

\end{document}